\documentclass[aps,prd,preprint,tightenlines,superscriptaddress,
amsfonts,amssymb,amsmath,byrevtex,showpacs,nofootinbib]{revtex4-2}
\usepackage[polish,english]{babel}
\usepackage[utf8]{inputenc}
\usepackage[T1]{fontenc}
\usepackage{latexsym}
\usepackage{graphicx}
\usepackage{geometry}
\usepackage{epstopdf}
\usepackage{subfig}
\usepackage{color}

\usepackage{xcolor}
\usepackage{pdfsync}
\usepackage{fancyhdr}
\usepackage{makeidx}
\usepackage[breaklinks,colorlinks,backref]{hyperref}
\hypersetup{
    colorlinks, 
    linktoc=all, 
    linkcolor=red,  
  citecolor=hyptxt,
  urlcolor=blue}
  \hypersetup{
  citebordercolor=,
  filebordercolor=green,
  linkbordercolor=blue
}
\definecolor{hyptxt}{rgb}{0.7, 0.4, 0.9}
\usepackage{bibtopic}
\newcommand{\nn}{\nonumber}
\newcommand{\bs}{\begin{subequations}}
\newcommand{\es}{\end{subequations} \noindent}
\newcommand{\bea}{\begin{eqnarray}}
\newcommand{\eea}{\end{eqnarray}}

\newcommand{\RNumb}{\mathbb{R}}

\newcommand{\dR}{\mathbb R}

\newcommand{\UnitOp}{\hat{1\kern-4.75pt 1}} 
\newcommand{\MatUnit}{1\kern-3pt 1} 
\newcommand{\Group}[1]{\textrm{#1}} 
\newcommand{\Aff}{{\textrm{Aff}(\mathbb{R})}} 
\newcommand{\Bra}[1]{\langle #1 \vert} 
\newcommand{\Ket}[1]{\vert #1 \rangle} 

\newcommand{\BraKet}[2]{\langle #1 \vert #2 \rangle} 

\newcommand{\Aver}[1]{\langle #1 \rangle} 

\newcommand{\Var}[1]{\mathrm{var}(#1)} 

\newcommand{\StateSpace}[1]{\mathcal #1} 
\newcommand{\Komentarz}[1]{} 
\begin{document}

\title{Quantum dynamics corresponding \\ to chaotic  BKL scenario}

\author{Andrzej G\'{o}\'{z}d\'{z}}
\email{andrzej.gozdz@umcs.lublin.pl}
\affiliation{Institute of Physics, Maria Curie-Sk{\l}odowska
University, pl.  Marii Curie-Sk{\l}odowskiej 1, 20-031 Lublin, Poland}

\author{Aleksandra P\c{e}drak} \email{aleksandra.pedrak@ncbj.gov.pl}
\affiliation{Department of Fundamental Research, National Centre for Nuclear
  Research, Pasteura 7, 02-093 Warszawa, Poland}

\author{W{\l}odzimierz Piechocki} \email{wlodzimierz.piechocki@ncbj.gov.pl}
\affiliation{Department of Fundamental Research, National Centre for Nuclear
  Research, Pasteura 7, 02-093 Warszawa, Poland}

\date{\today}

\begin{abstract}
We quantize the solution to the Belinski-Khalatnikov-Lifshitz (BKL) scenario
using the integral quantization method. Quantization smears the
gravitational singularity avoiding its localization in the configuration
space.  The latter is defined in terms of spatial and temporal coordinates,
which are treated on the same footing that enables respecting covariance of
general relativity. The relative quantum perturbations grow as the system
evolves towards the gravitational singularity. The quantum randomness
amplifies the deterministic classical chaos of the BKL scenario.
Additionally, our results suggest that the generic singularity of
general relativity can be avoided at quantum level giving support to the
expectation that quantum gravity has good chance to be a regular theory.
\end{abstract}


\maketitle

\tableofcontents

\section{Introduction}

The Belinski, Khalatnikov and Lifshitz (BKL) conjecture states that general
relativity includes the solution with generic gravitational singularity
\cite{BKL2,BKL3}. The evolution towards the BKL singularity, the so-called BKL
scenario, consists of the deterministic dynamics turning into stochastic process
near the generic singularity.   There are at least two fundamental
questions to be addressed in that context: What is the fate of the BKL chaos at quantum level?
Can the singularity be avoided in the corresponding quantum theory?

The evolution process presented in \cite{BKL2,BKL3} is complicated and difficult
to map into quantum evolution.  There exists well defined and comparatively
simple model of the BKL scenario \cite{bkl,Belinski:2014kba,book} that can be
used in the derivation of the BKL conjecture \cite{private}.  The model has been obtained
from the general model of the Bianchi IX spacetime for perfect fluid. The
equation of state of that fluid reads $p = k \varepsilon,~ 0\leq k < 1,$ where
$p$ and $\varepsilon$ denote the pressure and energy density of the fluid,
respectively. The case $k=1$ is excluded as it does not lead to the
oscillatory dynamics specific to the Bianchi IX model.  The massive BKL scenario model
has been obtained from that general Bianchi IX model by making the assumption (see
Eq. \!\eqref{order})
that in the dynamics, near the singularity, the anisotropy of space grows
without bound so that each of the so-called directional scale factors  oscillate,
but never crosses each other, and
evolves towards vanishing, i.e., singularity.  The resulting dynamics,
specified in the next section, is different from the commonly known mixmaster
dynamics \cite{Volodia,mix} in which one can divide the oscillatory evolution of the system
into  eras, each consisting of Kasner's epochs,  evolving towards the singularity.
The mixmaster model is the vacuum Bianchi IX model  and can serve to derive the BKL
conjecture as well \cite{private} so that we call it the vacuum model of the BKL scenario.
We have made the comparison of the dynamics of both massive and vacuum models
within dynamical systems method recently  \cite{ECz}. The dynamics of the massive BKL model depends
on the matter field implicitly via the directional scale factors which are effective ones,
see Eqs. (2.6), (2.7) and (2.24) in \cite{bkl}.
Also, due to the dependence of the general Bianchi IX model on matter components, it was
possible to obtain the asymptotic form (near the singularity) of the general dynamics \cite{bkl}.
That dynamics is mathematically simple enough to be solved analytically \cite{Piotr}.
The dynamics of the vacuum Bianchi IX model is the same faraway and near the singularity.
It is so complex that the model is non-integrable \cite{Conte}. Therefore, the massive
BKL model is both simpler for analysis than the vacuum BKL model, and better suitable to
describe the dynamics in the neighborhood of the cosmological singularity.

Recently, we have verified numerically that the classical dynamics underlying
the present paper leads to the gravitational singularity \cite{Nick1,Nick2},
and it is generically unstable turning into chaotic process near the
singularity \cite{Piotr}. These features are consistent with the original BKL
scenario \cite{BKL2,BKL3,book}.

As far as we are aware, there are not available any results concerning the
issue of the construction of quantum theory corresponding directly to the
original BKL scenario \cite{BKL2,BKL3}.  The existing results concern the Hamiltonian
framework, in terms of Ashtekar's variables, that is supposed to be convenient
to address the BKL scenario problem \cite{David1,David2}.  We have used the
homogeneous sector of that formalism to consider the possibility of existence
of classical end quantum spikes within that sector \cite{WPa,WPb}.  Recent article on
spacelike singularities of general relativity makes promise of coming back to
the issue of addressing the original BKL singularity problem within loop
quantum gravity \cite{Abhay}. We do not exclude the case
of joining that development in the future, but presently we rather prefer to
follow our programme.

Quantization of the dynamics presented in \cite{bkl} can be used in the
examination of the fate of the corresponding quantum dynamics. In fact, we have
already quantized that model with the conclusion that quantization of the
dynamics leads to avoiding gravitational singularity \cite{AWG,AW}. In these
papers, we have quantized Hamilton's dynamics  derived in \cite{Ewa}.
However, since quantization is known to be an ambiguous procedure, we have
decided to examine the robustness of these results by making use of a completely
different quantization method,  which is one of the goals of the present
paper.  That method, applied recently to the quantization of the
Schwarzschild spacetime \cite{AOW}, includes quantization of the temporal and
spatial variables on the same footing.  The rationale for such dealing is that
the distinction between time and space variables violates the general covariance
of arbitrary transformations of temporal and spatial coordinates.

The results of the present paper, in the context of resolving the cosmological
singularity, are similar to the results of the paper \cite{AOW} addressing the
issue of a singularity of an isolated object. In both cases
quantization smears the singularity avoiding its localization in configuration
space.  The issue of resolving the singularity within our two quite
different approaches will be further discussed in the last section.

The new phenomenon we deal with, in the present paper, is the fate of the
classical chaos at quantum level.  Our analysis shows that the
quantum randomness  turns deterministic classical chaos into stochastic
quantum chaos.

The paper is organized as follows: In Sec. II we recall the main results of the
paper \cite{Piotr} to have our paper self-contained. Sec. III presents the main
aspects of the coherent states quantization method adopted to our gravitational
system.  In Sec. IV we quantize the solution to the BKL scenario. Stochastic
aspects of quantum evolution are presented in Sec. V. We conclude in the last
section.  Appendix presents the essence of the coherent states quantization.

\noindent In the following we choose $\;G = c =1=\hbar\;$ except where otherwise
noted.

\section{Solution to the BKL scenario}

To have the paper self-contained, we recall in this section the main results of
Ref. \!\cite{Piotr} to be used later.

The massive model of the BKL scenario is defined to be \cite{bkl,book}
\begin{equation}\label{L1}
\frac{d^2 \ln a  }{d t^2} = \frac{b}{a}- a^2,~~~~\frac{d^2 \ln b
}{d t^2} = a^2 - \frac{b}{a} + \frac{c}{b},~~~~\frac{d^2 \ln c }{d
t^2} = a^2 - \frac{c}{b},
\end{equation}
subject to the constraint
\begin{equation}\label{L2}
\frac{d\ln a}{dt}\;\frac{d\ln b}{dt} + \frac{d\ln
a}{dt}\;\frac{d\ln c}{dt} + \frac{d\ln b}{dt}\;\frac{d\ln c}{dt} =
a^2 + \frac{b}{a} + \frac{c}{b} \, ,
\end{equation}
where $\,a=a(t)> 0,\, b=b(t)>0$ and $\,c=c(t)>0$ are the so-called
directional scale factors, while $t \in \dR$ is a monotonic function of proper time.

Eqs. \!\eqref{L1}--\eqref{L2} have been derived from the general
dynamics of the Bianchi IX model under the condition
that near the singularity the following strong inequalities  are satisfied \cite{bkl}
\begin{equation} \label{order}
a\gg b\gg c > 0 \, .
\end{equation}

It has been found in \cite{Piotr} that the analytical solutions to
Eqs. \!\eqref{L1}--\eqref{L2}, for $t>t_0$,  read
\begin{equation}\label{solution}
a(t)= \frac{3}{ t-t_0},~~~ b(t)= \frac{30}{
(t-t_0 )^{3}},~~~ c (t)= \frac{120}{( t-t_0 )^{5}} \, ,
\end{equation}
where   $ t - t_0  \neq 0$ and $t_0$ is an
arbitrary real number. Thus, the solutions are parameterized by the  number $t_0 \in \dR$.

The solution \eqref{solution} corresponds, for instance in the case $t > t_0$ and $t_0 < 0$, to the following choice of the initial
data
\begin{align}\label{initial}
\nonumber  a(0) &= - 3 \;t_0^{-1},&\dot{a} (0) &= - 3\; t_0^{-2}\,, \\
b(0) &= - 30\;t_0^{-3},& \dot{b}(0) &= - 90\; t_0^{-4}\,, \\
\nonumber  c(0) &= - 120\; t_0^{-5},& \dot{c}(0) &=  - 600\;
t_0^{-6}\,.
\end{align}

The stability analyses carried out in \cite{Piotr} have shown that the solution
\eqref{solution} is unstable against small perturbation. More precisely,
substituting the following functions
\bs\label{pert}
\begin{align}
& a (t) =3(t-t_0)^{-1}+\epsilon\alpha(t),\\
& b (t) =30(t-t_0)^{-3}+\epsilon\beta(t),\\
& c (t)=120(t-t_0)^{-5}+\epsilon\gamma(t) \, ,
\end{align}
\es
into \eqref{L1}--\eqref{L2} leads, in the first order in the small parameter
$\epsilon$, to the following solution of the resulting equations

\bs\label{asymp}
\begin{align}
\alpha (t)=&\exp(-\theta/2)\!\left[K_1\cos(\omega_1\theta\!+\!\varphi_1)\! +\!
K_2\cos(\omega_2\theta\!+\!\varphi_2)\right]
\!+\!K_3\exp(-2\theta) ,\\
\beta (t)=&\exp(-5\theta
/2)\left[\left(4+6\sqrt{6}\right)K_1\cos(\omega_1\theta+\varphi_1)\right.\nn\\&\left.+\left(4-6\sqrt{6}\right)K_2\cos(\omega_2\theta+\varphi_2)
\right]+30K_3\exp(-4\theta) ,\\
\gamma (t)=&-4\exp(-9\theta
/2)\left[\left(26+9\sqrt{6}\right)K_1\cos(\omega_1\theta+\varphi_1)\right.\nn\\&\left.+\left(26-9\sqrt{6}\right)K_2\cos(\omega_2\theta+\varphi_2)
\right]+200 K_3\exp(-6\theta) \, ,
\end{align}
\es
where $\theta=\ln (t-t_0)$.  The two frequencies read
\begin{equation}\label{freqs}
\omega_1 = \frac{1}{2}\sqrt{95-24\sqrt{6}},\qquad\omega_2 =
\frac{1}{2}\sqrt{95+24\sqrt{6}} \, ,
\end{equation}
where $K_1,K_2,K_3, \varphi_1,$ and $\varphi_2$ are constants.\\

The manifold $\mathcal{M}$ defined by
$\{ K_1, K_2, K_3, \varphi_1, \varphi_2 \}$ is a submanifold of $\dR^5$. The
solution defined by \eqref{pert} and \eqref{asymp} corresponds to the choice of
the set of the initial data $\mathcal{N}$ that is a small neighborhood of the
initial data \eqref{initial}. $\mathcal{N}$ is a submanifold of $\dR^5$ as
\eqref{initial} defines five independent constants due to the constraint
\eqref{L2}. Therefore, it is clear that \eqref{asymp} presents a generic
solution as the measures of both $\mathcal{M}$ and $\mathcal{N}$ are
nonzero. The exact solution \eqref{solution} alone is of zero-measure in the
space of all possible solutions to Eqs. \!\eqref{L1}--\eqref{L2}.

The relative perturbations $\alpha/a, \beta/b,$ and $\gamma/c$ grow
proportionally as $\exp(\frac{1}{2}\theta)$. The multiplier $1/2$ plays the role
of a Lyapunov exponent, describing the rate of their divergence. Since it is
positive, the evolution of the system towards the gravitational singularity
($\theta \rightarrow + \infty$) is chaotic. The transition into the chaos occurs
if the evolution begins with the initial data which belong, for instance, to the
neighbourhood of the conditions \eqref{initial}.

The original BKL scenario \cite{BKL2,BKL3} is known to  enter the chaotic phase
near the singularity. Its vacuum model, the mixmaster universe, has been proved to include
the chaotic dynamics  \cite{st1,st2,st3,st4,st5}. Its massive model \cite{Piotr}, underlying the present paper,
has never been examined earlier in the context of stochasticity. Finding that its dynamics is chaotic,
opens the door for studies of that issue at quantum level, which is the main subject of our article.
The main difference between both BKL models (in terms of physics) is that the latter is more realistic
near the singularity as it includes effectively some contribution from matter field.

\section{Affine coherent states quantization}

We propose to quantize the classical BKL scenario by using the integral
quantization called the affine coherent states quantization, see Appendix.
We have recently applied this approach in the context of cosmology \cite{AWG,AW}
and astrophysics \cite{AOW,WPTS}.

In general relativity time and position in space are treated on the same level,
however, in quantum mechanics time is not considered to be a quantum observable,
but  a parameter enumerating events.  In this paper, we treat time and position
on the same footing in the quantum description. They are related to operators
obtained by the affine coherent states quantization. This idea requires
introducing the notion of an extended classical configuration space by including
time as an additional coordinate. The correspondence between the classical time
and position is done by comparing their classical values with expectation values
of their quantum counterparts.

In the following, we extend the method of quantization used in \cite{AWG,AW,AOW}.
In the present description of the BKL scenario, the
Hilbert space has to be extended to the carrier space of infinite
dimensional unitary irreducible representation of the direct product of three
affine groups with additional constraints determining a model of physical
time. In this approach there is no distinction between kinematical and
dynamical Hilbert spaces; we construct the state space of the system
with time treated on the same footing as other observables. In addition, we
obtain quantum states evolving similarly to
classical solutions. It is achieved by using the idea of the correspondence
principle between quantum mechanics and its classical approximation.

To begin with, we introduce two configuration spaces defined as follows: the
classical gravitational configuration space $T_{BKL}$
\begin{equation}\label{a10}
T_{BKL} := \{ (t,a,b,c): (t,a,b,c)  \in \RNumb \times \RNumb_+^3  \} \, ,
\end{equation}
where $\RNumb_+  = (0,+\infty)$, and the affine configuration space $T$
is defined as
\begin{equation}\label{a1}
T = \{\xi= (\xi_1,\xi_2,\xi_3,\xi_4,\xi_5,\xi_6):
\xi \in (\RNumb \times \RNumb_+) \times (\RNumb \times \RNumb_+)
\times (\RNumb \times \RNumb_+) \} \, ,
\end{equation}
where every pair $(\xi_k,\xi_{k+1})$, ($k=1,3,5$), parameterizes the affine
group $\Aff$.

The variables with even indices correspond to the scale factors
$\xi_2=a,\,\xi_4=b,\,\xi_6=c$. Because $a,b,c >0$ and
$\xi_1,\xi_3,\xi_5 \in \RNumb$,  the configuration space $T$ parametrizes the
simple product of 3 affine groups $\Aff \times \Aff \times \Aff =: G$ to be used
in the affine quantization.

As the observational data are parameterized by a single time parameter, the
variables $\xi_1,\,\xi_3,\,\xi_5$ should be mapped onto a single variable
representing time.

The affine group $\Aff$ is known to have two nontrivial unitary irreducible
representations in the Hilbert space $\mathcal{H}_x := L^2(\dR_+,d\nu(x))$,
where $d\nu(x):= dx/x$. We choose the one defined as follows  (the second
representation would give exactly the same results):
\begin{equation}\label{a2}
  U(\xi_k,\xi_{k+1})\Psi(x) = e^{i\xi_k x}\Psi(\xi_{k+1} x),
\end{equation}
where, $k=1,3,5$ and $\BraKet{x}{\Psi} =: \Psi (x)\in \mathcal{H}_x$. The action
\eqref{a2} corresponds to the standard parametrization of the affine group
$\Aff$ defined by the multiplication law
\begin{equation}\label{a3}
(\xi_k,\xi_{k+1}) \cdot (\xi_k^\prime,\xi_{k+1}^\prime)
:= (\xi_k + \xi_{k+1} \xi_k^\prime, \xi_{k+1} \xi_{k+1}^\prime) \in \Aff \, .
\end{equation}
The left invariant measure on the group $\Aff$ reads
\begin{equation}\label{a4}
  d\mu (\xi_k,\xi_{k+1}) := d\xi_k\;\frac{d\xi_{k+1}}{\xi_{k+1}^2},
\end{equation}
and the corresponding invariant integration over the affine group is defined to
be
\begin{equation}\label{a5}
  \int_{\Aff} d\mu (\xi_k,\xi_{k+1}) :=
  \frac{1}{2\pi}\int_{-\infty}^\infty d\xi_k
  \int_0^\infty d\xi_{k+1} /\xi_{k+1}^2 \, .
\end{equation}
It is clear that the  direct product of three affine groups $G$ has the
unitary irreducible representation in the following Hilbert space
$\mathcal{H} = \mathcal{H}_{x_1}\otimes\mathcal{H}_{x_2}\otimes\mathcal{H}_{x_3}
= L^2(\RNumb_+^3, d\nu(x_1,x_2,x_3))$, where
$d\nu(x_1,x_2,x_3) = d\nu(x_1)d\nu(x_2)d\nu(x_3)$. It enables defining in
$\mathcal{H}$ the continuous family of affine coherent states
$\BraKet{x_1,x_2,x_3}{\xi_1,\xi_2;\xi_3,\xi_4;\xi_5,\xi_6} :=
\BraKet{x_1}{\xi_1,\xi_2} \BraKet{x_2}{\xi_3,\xi_4} \BraKet{x_3}{\xi_5,\xi_6}$,
as follows
\begin{equation}\label{a7}
\mathcal{H} \ni \BraKet{x_1,x_2,x_3}{\xi_1,\xi_2;\xi_3,\xi_4;\xi_5,\xi_6}
 := U (\xi)\Phi_0(x_1,x_2,x_3) \, ,
\end{equation}
where $U(\xi):= U (\xi_1,\xi_2) U (\xi_3,\xi_4)U (\xi_5,\xi_6)$, and
$\Ket{\xi_1,\xi_2;\xi_3,\xi_4;\xi_5,\xi_6}:=\Ket{\xi_1,\xi_2} \Ket{\xi_3,\xi_4}
\Ket{\xi_5,\xi_6}$  and where
\begin{equation}\label{a7A}
\mathcal{H} \ni \Phi_0(x_1,x_2,x_3)= \Phi_1 (x_1)\Phi_2 (x_2)\Phi_3 (x_3)\, .
\end{equation}
In \eqref{a7A} the vectors $\Phi_k (x_k) \in L^2(\RNumb_+, d\nu(x_k))$,
$k=1,2,3$ are the so-called fiducial vectors. They are required to satisfy the
two conditions
\begin{equation}\label{aa7A0}
  \int_0^\infty \frac{dx}{x}|\Phi_k(x)|^2=1 \, ,
\end{equation}
and
\begin{equation}\label{aa7A}
 A_{\phi_l} := \int_0^\infty \frac{dx}{x^2}|\Phi_k(x)|^2 < \infty \, .
\end{equation}
where $l=1$ for $k=1$, $l=3$ for $k=2$ and $l=5$ for $k=3$.
The fiducial vectors are the free ``parameters'' of this quantization scheme.

Finally, we have
\begin{eqnarray}
 \nonumber  U (\xi_1,\xi_2,\xi_3,\xi_4,\xi_5,\xi_6) \Phi_0(x_1,x_2,x_3)
 &=&  \nonumber \\
   U (\xi_1, \xi_2) U (\xi_3, \xi_4)  U (\xi_5, \xi_6)\Phi_1
(x_1)\Phi_2 (x_2)\Phi_3 (x_3) &=& \nonumber \\
\label{a6}  e^{i(\xi_1 x_1 +\xi_3 x_2 +\xi_5 x_3 )}
\Phi_1 (\xi_2 x_1) \Phi_2(\xi_4 x_2) \Phi_3 (\xi_6 x_3) \, .
\end{eqnarray}
The irreducibility of the representation leads to the resolution of the unity in
$\mathcal{H}$ as follows
\begin{eqnarray}
\label{a8}
&&\frac{1}{A_\phi} \int_\Group{G} d\mu(\xi) \Ket{\xi}\Bra{\xi}
:=  \bigotimes_{k=1,3,5} \frac{1}{A_{\phi_k}}
\int_{\Aff} d\mu(\xi_k,\xi_{k+1})
\Ket{\xi_k,\xi_{k+1}}\Bra{\xi_k,\xi_{k+1}} \nonumber \\
&& =\frac{1}{A_{\Phi_1}A_{\Phi_3}A_{\Phi_5}}
\int_\Aff d\mu(\xi_1,\xi_2) \Ket{\xi_1,\xi_2} \Bra{\xi_1,\xi_2}
\otimes
\int_{\Aff} d\mu(\xi_3,\xi_4) \Ket{\xi_3,\xi_4} \Bra{\xi_3,\xi_4}  \nonumber \\
&& \otimes \int_{\Aff} d\mu(\xi_5,\xi_6) \Ket{\xi_5,\xi_6} \Bra{\xi_5,\xi_6}
= \UnitOp_1\otimes \UnitOp_2\otimes \UnitOp_3 = \UnitOp \, ,
\end{eqnarray}
where $d\mu(\xi)=\Pi_{k=1,3,5} d\mu(\xi_k,\xi_{k+1})$.

The resolution \eqref{a8} can be used for mapping a classical observable
$f: T \rightarrow \RNumb$ into an operator
$\hat{f}: \mathcal{H} \rightarrow \mathcal{H}$ as follows \cite{AWG,AW,AWT,AOW}
\begin{eqnarray}
\label{a88}
&&  \hat{f} := \frac{1}{A_\phi} \int_{\Group{G}} d\mu(\xi)
\Ket{\xi} f(\xi)\Bra{\xi} \nonumber \\
&& =\frac{1}{A_{\Phi_1}A_{\Phi_3}A_{\Phi_5}}\int_\Aff
d\mu(\xi_1,\xi_2) \int_\Aff d\mu(\xi_3,\xi_4) \int_\Aff d\mu(\xi_5,\xi_6)
\nonumber \\
&& \Ket{\xi_1,\xi_2;\xi_3,\xi_4;\xi_5,\xi_6}
f(\xi_1,\xi_2;\xi_3,\xi_4;\xi_5,\xi_6)
\Bra{\xi_1,\xi_2;\xi_3,\xi_4;\xi_5,\xi_6} \, ,
\end{eqnarray}
where $A_\Phi:= A_{\Phi_1}A_{\Phi_3}A_{\Phi_5}$.

Here, we recall two standard characteristics of quantum observables:
(i) expectation values and variances of quantum observables are the most important
characteristics which allow to compare quantum and classical worlds, and
(ii) expectation values of quantum observables correspond to classical values of
measured quantities and their variances describe quantum smearing of these
observables.

A general form of expectation value of the observable $\hat{f}$ obtained from
the classical function $f$, while the quantum system is in the state
$|\Psi \rangle$, reads
\begin{equation}
\label{eq:ExpVal}
\Aver{\hat{f};\Psi}:=\Bra{\Psi}\hat{f}\Ket{\Psi}
= \frac{1}{A_\phi} \int_\Group{G} d\mu(\xi) f(\xi) |\BraKet{\xi}{\Psi}|^2 \, .
\end{equation}
The variance of an observable $\hat{f}$ defined as
$\Var{\hat{f};\Psi}:= \Aver{(\hat{f} -\Aver{\hat{f};\Psi})^2, \Psi}$ is more
difficult for calculations because it requires $\Aver{\hat{f}^2;\Psi}$ which
involves an overlap between the coherent states, and usually depends on the
fiducial vector explicitly:
\begin{equation}
\label{eq:VarVal}
\Aver{(\hat{f})^2;\Psi}:=\Bra{\Psi} (\hat{f})^2 \Ket{\Psi}
= \frac{1}{A_\phi} \int_\Group{G} d\mu(\xi)
\frac{1}{A_\phi} \int_\Group{G} d\mu(\xi')
\BraKet{\Psi}{\xi}f(\xi)  \BraKet{\xi}{\xi'} f(\xi') \BraKet{\xi'}{\Psi} \, .
\end{equation}

The variance $\Var{\hat{f};\Psi} $ can be rewritten as
\begin{equation}
\label{eq:VarVal2}
\Var{\hat{f};\Psi} =
\Aver{\hat{f}^2;\Psi} - \Aver{\hat{f};\Psi}^2 \, .
\end{equation}
The  important quantum observables correspond to the variables of the
configuration space  \eqref{a1}.
These elementary variables $\xi_k$ ($k=1,2,\dots,6$), can be mapped into the
quantum operators as follows
\begin{equation}
\label{eq:ElementQO}
\hat{\xi}_k
= \frac{1}{A_\Phi} \int_{\Group{G}} d\mu(\xi)\Ket{\xi} \xi_k\Bra{\xi} \, .
\end{equation}
For every $k$ the above equality \eqref{eq:ElementQO} reduces to integration
over a single affine group. The other integrations give the unit operators
in two remaining spaces  $\mathcal{H}_{x_l}$, $l \not= k$. For example,
\begin{equation}
\hat{\xi}_2
= \frac{1}{A_{\Phi_1}} \int_{\Aff} d\mu(\xi_1,\xi_2)
\Ket{\xi_1,\xi_2} \xi_2\Bra{\xi_1,\xi_2}
\otimes \UnitOp_{x_1} \otimes \UnitOp_{x_3} \label{eq:ElementQO2} \ .
\end{equation}

To deal with a single time variable at quantum level, one needs to choose a
model of time in the configuration space $T$, defined by \eqref{a1}.  In
general, it can be introduced either as a real function or distribution,
$\mathcal{T}: T \to \RNumb$. Its quantization leads to the time operator
$\hat{\mathcal{T}}$.  However, we can impose the appropriate constraints to have
the common time variable for all three operators $\hat{\xi}_1, \hat{\xi}_3$, and
$\hat{\xi}_5$.  In this paper we realise that option assuming that the only
allowed quantum states $\Psi$ of our BKL system are the states which satisfy the
condition
\begin{equation}
\label{eq:TimeCond}
\Bra{\Psi}\hat{\xi}_1\Ket{\Psi}=\Bra{\Psi}\hat{\xi}_3\Ket{\Psi}
=\Bra{\Psi}\hat{\xi}_5\Ket{\Psi} \ ,
\end{equation}
which is consistent with the choice of the configuration space in the form
\eqref{a10}.
It means, we require the same expectation values for all three operators
which represent three ``times'' related to appropriate quantum
observables
$\hat{a} = \hat{\xi}_2, \hat{b} = \hat{\xi}_4$, and $\hat{c}
= \hat{\xi}_6$.

\section{Quantization of the solution to  BKL scenario}

The above quantization scheme can be now applied to the solutions \eqref{asymp}
of the BKL scenario\footnote{In what follows we denote by
  $\tilde{a}, \tilde{b}$ and $\tilde{c}$ the solution \eqref{solution}.}
\begin{equation}\label{a12}
  a(t) = \tilde{a}(t) + \epsilon \alpha (t),~~~~b(t) = \tilde{b}(t) + \epsilon
  \beta (t),~~~~c(t) = \tilde{c}(t) + \epsilon \gamma (t) \, ,
\end{equation}
ascribing to them appropriate quantum states and the corresponding operators.

In quantum mechanics, contrary to classical one, one needs two kinds of objects
to describe physical world. These are quantum observables represented by either
appropriate operators or operator valued measures, and quantum states being the
vectors in a Hilbert space or the so called density operators. In the classical
mechanics the functions on either configuration or phase space are at the same
time states and observables.

To quantize solutions of the BKL scenario, we already have the elementary
observables $\hat{\xi}_k$. However, we also have to find the appropriate family of
states related to the solutions \eqref{a12}. This family of states has to
reproduce the classical solutions \eqref{a12} by comparing them with expectation
values of the corresponding observables.

The classical solutions are represented by three time dependent functions.
In the configuration space $T$ we have 6 variable, where 3 of them
$\xi_1,\xi_3,\xi_5$ represent the time in the state space which satisfies the
condition \eqref{eq:TimeCond}.  As is it was mentioned above, the classical
observables should be related to their quantum counterparts by the
corresponding expectation values.  This idea leads directly to the conditions
for a family of states
$\{\Psi_\eta(x_1,x_2,x_3)=\BraKet{x_1,x_2,x_3}{\Psi_\eta}, \eta \in \dR^s
\}$ parameterized by a set of evolution parameters
 $\eta=(\eta_1,\eta_2,\dots\eta_s)$ enumerating the set of trial functions.

We require the states $\Ket{\Psi_\eta}$  to satisfy the following conditions
\cite{AOW}:

\begin{eqnarray}
&& \Bra{\Psi_\eta} \hat{\xi}_k \Ket{\Psi_\eta}=t ,\quad k=1,3,5
\label{eq:EqMotions1} \\
&& \Bra{\Psi_\eta} \hat{\xi}_2 \Ket{\Psi_\eta}=a(t) \ ,
\label{eq:EqMotions2} \\
&& \Bra{\Psi_\eta} \hat{\xi}_4 \Ket{\Psi_\eta}=b(t) \ ,
\label{eq:EqMotions3} \\
&& \Bra{\Psi_\eta} \hat{\xi}_6 \Ket{\Psi_\eta}=c(t) \ .
\label{eq:EqMotions4}
\end{eqnarray}

The equations \eqref{eq:EqMotions1} represent the single time constraint
\eqref{eq:TimeCond}.
The parameter $\eta$ labels the family of states to be found, and it should be a function of $t$
as the r.h.s. of (\ref{eq:EqMotions1})--(\ref{eq:EqMotions4}) depends on $t$.
The solution of Eqs.~(\ref{eq:EqMotions1})--(\ref{eq:EqMotions4}) allows to construct
the vector state dependent on classical time,  $\Ket{\Psi_{\eta(t)}} \in \StateSpace{H}$,
in our Hilbert space.

In this way we relate the quantum dynamics to the classical one.  Obviously,
there may exist more than one family of states satisfying the above equations of
motion.

In what follows, we determine the states $\Ket{\Psi_{\eta(t)}}$ satisfying the
conditions \eqref{eq:EqMotions1}--\eqref{eq:EqMotions4}.  This will enable
examination of the issue of the fate of the gravitational singularity and chaos
of the BKL scenario at the quantum level.

\subsection{Evolving wave packets}

In our paper we consider two kinds of wave packets satisfying the conditions
\eqref{eq:EqMotions1}--\eqref{eq:EqMotions4}.
The first kind are the affine coherent states themselves. The second type is a
set of modified ``exponential'' wave packets, which  represent a dense set of
states in the Hilbert space $\mathcal{H}$.

\subsubsection{Coherent states  and  expectation values}

One can verify that considered coherent states generated by a
single affine group satisfy, due to the results of the recent paper \cite{AOW},
the following equations
\begin{equation}\label{a17}
\Bra{\xi_k,\xi_{k+1}}\hat{\xi}_l \Ket{\xi_k,\xi_{k+1}}= \xi_l,
\text{ where } l=k,k+1; \ k=1,3,5 \, ,
\end{equation}
where the operators  $\hat{\xi}_l $  are defined to be
\begin{equation}\label{ex1}
\hat{\xi}_l := \frac{1}{2\pi} \frac{1}{A_{\Phi_k}}
\int_\RNumb d\xi_k \int_{\RNumb_+} \frac{d\xi_{k+1}}{\xi_{k+1}^2}\;
\Ket{\xi_k,\xi_{k+1}} \xi_l \Bra{\xi_k,\xi_{k+1}} \, ,
\end{equation}
and where $l=k,k+1$, $k=1,3,5$.

The conditions \eqref{a17} are the consistency conditions between the affine
group parametrization and the configuration space of the quantized physical
system \cite{AOW}.

This implies that the coherent states generated by the product of three affine
groups also satisfy the consistency condition
\begin{equation}\label{a17A}
\Bra{\xi}\hat{\xi}_l \Ket{\xi}
= \frac{1}{A_\phi} \int_G d\mu(\xi')
\BraKet{\xi}{\xi'} \xi_l' \BraKet{\xi'}{\xi}= \xi_l,
\text{ where } l=1,2,\dots,6 \, .
\end{equation}

The consistency conditions coincide with an idea that the coherent state
$\Ket{\xi_1,\xi_2,\xi_3,\xi_4,\xi_5,\xi_6}$ represents a state localized at
the point $\{\xi_1,\xi_2,\xi_3,\xi_4,\xi_5,\xi_6\}$ of the affine configuration
space and at the same time in the spacetime.

Therefore, the coherent states
\begin{equation}
\label{eq:CSWavePack3}
\Ket{CS_\epsilon;t}:=
 \Ket{t, \tilde{a}(t)+\epsilon \alpha(t);\,
  t, \tilde{b}(t)+\epsilon\beta(t);\, t, \tilde{c}(t)+\epsilon\gamma(t)}
\end{equation}

satisfy the equations of motions \eqref{eq:EqMotions1}--\eqref{eq:EqMotions4}.
In such case, we propose to use a one dimensional parameter $\eta$ and identify it
with the classical time $t$, i.e., the classical  time is a label of the
evolving family of quantum states.

Realization of \eqref{eq:CSWavePack3} as a wave packet constructed in the
  space of square integrable functions $L^2(\RNumb_+^3,d\nu(x_1,x_2,x_3))$ reads
\begin{eqnarray}
\label{eq:CSWavePack2}
&& \Psi_{CS_\epsilon}(t,x_1,x_2,x_3)
=\BraKet{x_1,x_2,x_3}{CS_\epsilon;t} \nonumber \\
&&=e^{it(x_1+x_2+x_3)} \Phi_1(a(t) x_1)\Phi_2(b(t) x_2)\Phi_3 (c(t) x_3)\, .
\end{eqnarray}

 Using equation \eqref{a17A} and the fact that the vector \eqref{eq:CSWavePack3}
factorises
\begin{equation}
\Ket{CS_\epsilon;t}=
 \Ket{t, \tilde{a}(t)+\epsilon \alpha(t)}\Ket{
  t, \tilde{b}(t)+\epsilon\beta(t)}\Ket{ t, \tilde{c}(t)+\epsilon\gamma(t)} \, ,
\end{equation}
we can compute the expectation value of the volume  operator $\hat{V}$,
where $V := \xi_2\xi_4\xi_6$, as follows
\begin{equation}
\label{eq:ExpABC_CS}
\Bra{CS_\epsilon;t}\hat{V}\Ket{CS_\epsilon;t}=
(\tilde{a}(t)+\epsilon \alpha(t))\,
(\tilde{b}(t)+\epsilon\beta(t))\,
(\tilde{c}(t)+\epsilon\gamma(t)) \, .
\end{equation}

\subsubsection{The modified exponential packet  and  expectation values}

Let us consider the set of Gaussian distribution wave packets (with modified
exponential part)
\begin{equation}
\label{eq:EWavePack}
\Psi_n(x; \tau,\gamma)=Nx^n \exp\left[ i\tau x-\frac{\gamma^2 x^2}{2}\right]\, ,
\quad N^2=\frac{2\gamma^n}{(n-1)!} \, ,
\end{equation}

which according to \cite{AOW} is dense in $L^2(\RNumb_+, d\nu(x))$.

The expectation values and
variances of the operators $\hat{\xi}_k$ and $\hat{\xi}_{k+1}$ have the
following values
\begin{eqnarray}
&&\Bra{\Psi_n}\hat{\xi}_k\Ket{\Psi_n}=\tau,~~~~k= 1,3,5 \, ,\\
&&
\label{eq:ExpXik1}
\Bra{\Psi_n}\hat{\xi}_{k+1}\Ket{\Psi_n}=\frac{1}{A_\Phi}\frac{\Gamma\left(n-\frac{1}{2}\right)}{(n-1)!}\gamma \, ,\\
&&\Var{\hat{\xi}_k;\Psi_n}=\frac{4n-3}{4(n-1)}\gamma^2 \, ,\\
&&\Var{\hat{\xi}_{k+1};\Psi_n}=\frac{1}{A_\Phi^2}\left(\frac{1}{n-1}-\frac{\Gamma\left(n-\frac{1}{2}\right)^2}{(n-1)!^2}\right)\gamma^2 \, .
\end{eqnarray}
 In this case,
the evolution parameters $\eta$ consists of $\eta_1=\tau$ and $\eta_2=\gamma$.

In the space $L^2(\RNumb_+^3,d\nu(x_1,x_2,x_3))$ we take the corresponding  wave
packets in the form
\begin{equation}
\label{eq:EWavePackAff3}
\Psi_{n_1,n_3,n_5}(x_1,x_2,x_3;\,
\tau_1,\tau_3,\tau_5,\gamma_1,\gamma_3,\gamma_5)
=
\Psi_{n_1}(x_1; \tau_1,\gamma_1)\Psi_{n_3}(x_2; \tau_3,\gamma_3)\Psi_{n_5}(x_3;
\tau_5,\gamma_5) \, .
\end{equation}
To meet the properties \eqref{eq:EqMotions1}--\eqref{eq:EqMotions4} for the wave
packets $\Psi_{n_1,n_3,n_5}$, we choose the parameters $\tau_k$ and
$\gamma_k$ as follows
\begin{eqnarray}
&&\tau_1=\tau_3=\tau_5=t \, ,
\label{eq:EWavePackCond1}
\\
&&\gamma_k=A_{\Phi_k}\frac{(n_k-1)!}{\Gamma\left(n_k-\frac{1}{2}\right)}\cdot
f_k(t)\, ,
\quad k=1,3,5 \, ,
\label{eq:EWavePackCond4}
\end{eqnarray}
where
\begin{equation}
f_k(t)=\left\{
\begin{array}{ll}
\tilde{a}(t)+\epsilon \alpha(t)\, ,&k=1\\
\tilde{b}(t)+\epsilon\beta (t)\, ,&k=3\\
\tilde{c}(t)+\epsilon\gamma(t)\, ,&k=5
\end{array}
\right. \, .
\end{equation}

Using similar technique as in case equation \eqref{eq:ExpABC_CS}, one
can calculate the
expectation value of the volume operator in the state \eqref{eq:EWavePackAff3}
and one gets
\begin{equation}
\Bra{\Psi_{n_1,n_3,n_5}}\hat{V}\Ket{\Psi_{n_1,n_3,n_5}}=f_1(t)f_3(t)f_5(t) \, ,
\end{equation}
which coincides with the result obtained within coherent states method \eqref{eq:ExpABC_CS}.

It is clear that the expectation value of the volume operator converges very fast to zero as
$t \rightarrow \infty$.

\subsection{Variances in the Hilbert space $\mathcal{H}$}

The wave packets obtained above follow the classical solutions.
However, in quantum mechanics the observables which do not commute with time
and position operators fluctuate at every spacetime point. This smearing of
quantum observables is determined by the Heisenberg uncertainty principle.
The most important ingredients of it are variances of the corresponding observables.
In the following we perform analysis of the variances of our elementary variables.

\subsubsection{Using coherent states}

The variance of the operators $\hat{\xi}_k$, $\hat{\xi}_{k+1}$,
 $k=1,3,5$ in coherent states \eqref{eq:CSWavePack3} read
\begin{eqnarray}
&&\label{vc1} \Var{\hat{\xi}_k;\Ket{CS_\epsilon;t}}
=\Aver{\hat{\xi}^2_k}_0 f_k(t)^2 \, , \\
&&\label{vc2}
\Var{\hat{\xi}_{k+1};\Ket{CS_\epsilon;t}}
=\left(\Aver{\hat{\xi}^2_{k+1}}_0-1\right) f_k(t)^2 \, ,
\end{eqnarray}
where
\begin{eqnarray}\label{ucs1}
&&\Aver{\hat{\xi}^2_i}_0=\Bra{0,1}\hat{\xi}^2_i\Ket{0,1},
\quad i=1,2,\dots 6 \, ,
\end{eqnarray}
and where $\Aver{\hat{\xi}^2_i}_0$ is a constant which depends on
the  choice of the fiducial vector.

The variance of the volume operator is found to be
\begin{equation}
\Var{\hat{V}; \Ket{CS_\epsilon;t}}=
\left[\prod_{k=2,4,6}\Aver{\hat{\xi}^2_k}_0-1\right]
f_1(t)^2 f_3(t)^2 f_5(t)^2 \, .
\end{equation}

For more details concerning the r.h.s. of \eqref{ucs1} see App. D of \cite{AOW}.

\subsubsection{Using exponential wave packet}

The corresponding results for the wave packets \eqref{eq:EWavePackAff3}, under
the conditions \eqref{eq:EWavePackCond1}-\eqref{eq:EWavePackCond4}, read
\begin{eqnarray}
&&\label{vp1} \Var{\hat{\xi}_k; \Psi_{n_1,n_3,n_5}}=\mathcal{A}_k f_k(t)^2 \, ,\\
&&\label{vp2} \Var{\hat{\xi}_{k+1}; \Psi_{n_1,n_3,n_5}}=\mathcal{B}_k f_k(t)^2 \, ,
\end{eqnarray}
where
\begin{eqnarray}
&&
\mathcal{A}_k=
A^2_{\Phi_k}\frac{(4n_k-3)(n_k-1)!(n_k-2)!}{4\Gamma
  \left(n_k-\frac{1}{2}\right)^2}
\, ,\\
&& \mathcal{B}_k=\frac{(n_k-1)!(n_k-2)!}{\Gamma\left(n_k-\frac{1}{2}\right)^2}-1
\, .
\end{eqnarray}

The variance of the volume operator has the form
\begin{equation}
\Var{\hat{V};\Psi_{n_1,n_3,n_5}}=
\left[\prod_{k=2,4,6}\frac{(n_k-1)!(n_k-2)!}{\Gamma\left(n_k-\frac{1}{2}\right)}-1\right]f_1(t)^2 f_3(t)^2 f_5(t)^2 \, .
\end{equation}

These results show that all positions of our system in time and space are smeared
owing to nonzero variances. It is an important fact about possibility of avoiding singularities
in this dynamics.

\section{Stochastic aspects of quantum evolution}

The results of recent paper \cite{Piotr} give strong support to the expectation
that near the generic gravitational singularity the evolution becomes
chaotic. In what follows we examine that fundamental property of the BKL
scenario at quantum level.

To make direct comparison with the results of \cite{Piotr}, we split the
variances \eqref{vp2} into the contributions from unperturbed and perturbed
states. We have:
\begin{eqnarray}
 f_2(t)^2  &=& (\tilde{a}(t) + \epsilon \alpha (t))^2 = \tilde{a}(t)^2 + 2\epsilon \tilde{a} (t) \alpha (t) + \epsilon^2 \alpha (t)^2 \simeq
 \tilde{a}(t)^2 + 2\epsilon \tilde{a} (t) \alpha (t)\, ,   \label{vv1}\\
 f_4(t)^2  &=& (\tilde{b} (t) + \epsilon \beta (t))^2 = \tilde{b}(t)^2 + 2\epsilon \tilde{b} (t) \beta (t) + \epsilon^2 \beta (t)^2 \simeq
 \tilde{b}(t)^2 + 2\epsilon \tilde{b} (t) \beta (t) \, ,   \label{vv2} \\
 f_6(t)^2 &=& (\tilde{c} (t)+ \epsilon \gamma (t))^2 = \tilde{c}(t)^2 + 2\epsilon \tilde{c} (t) \gamma (t) + \epsilon^2 \gamma (t)^2 \simeq
 \tilde{c}(t)^2 + 2\epsilon \tilde{c} (t) \gamma (t)    \, .    \label{vv3}
\end{eqnarray}

The corresponding dimensionless functions describing relative quantum
perturbations are defined to be
\begin{equation}\label{vvv}
\kappa_k := \frac{\Var{\hat{\xi}_{k}; \Psi_{pert}}
  - \Var{\hat{\xi}_{k}; \Psi_{unpert}}}
{\Var{\hat{\xi}_{k}; \Psi_{unpert}}},~~~~~k = 2, 4, 6 \, ,
\end{equation}
where $\Psi_{pert}$ and $\Psi_{unpert}$ denote perturbed and unperturbed
wave packets, respectively.

The explicit form of \eqref{vvv}, up to the 1-st order in $\epsilon$, reads:
\begin{eqnarray}
\label{rqp1} \kappa_a (t ) := \kappa_2 (t ) &=&
\frac{2\epsilon \tilde{a}(t) \alpha(t)}{\tilde{a}(t)^2}
= 2\epsilon \frac{\alpha(t)}{\tilde{a}(t)}  \, ,\\
\label{rqp2} \kappa_b (t):= \kappa_4 (t) &=&
\frac{2\epsilon \tilde{b}(t) \beta(t)}{\tilde{b}(t)^2}
=  2\epsilon \frac{\beta(t)}{\tilde{b}(t)} \, , \\
\label{rqp3}   \kappa_c (t):= \kappa_6 (t) &=&
\frac{2\epsilon \tilde{c}(t) \gamma (t)}{\tilde{c}(t)^2}
= 2\epsilon \frac{\gamma(t)}{\tilde{c}(t)}\, .
\end{eqnarray}

It is clear that the relative perturbations \eqref{rqp1}--\eqref{rqp3} are the same for
the coherent states and the exponential wave packets.

\begin{figure}
\begin{center}
\includegraphics[width=0.7\textwidth]{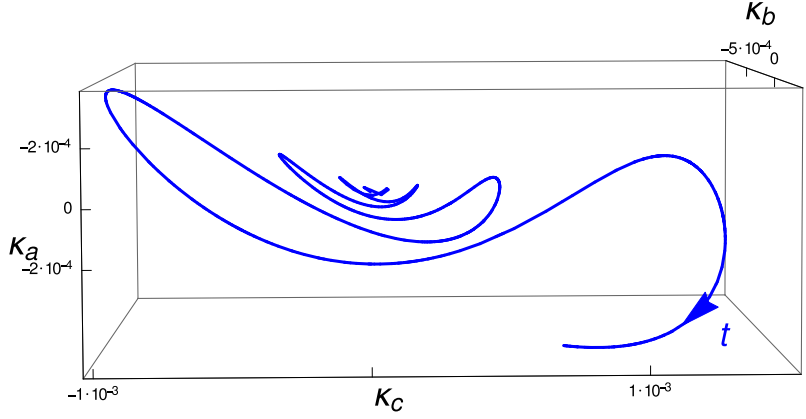}
\caption{\label{plot:3D}
The $t$ dependence of quantum perturbation defined by \eqref{rqp1}-\eqref{rqp3} for
$K_1=K_2=0.01$, $K_3=0$, $\phi_1=\phi_2=0$, $t_0 < 0$, $\epsilon=0.01$. The plot presents the parametric
curve $\{\kappa_a(t),\kappa_b(t),\kappa_c(t)\}$, where $t-t_0 \in(0.01, 35)$.
}
\end{center}
\end{figure}

\begin{figure}
\begin{center}
\includegraphics[width=0.47\textwidth]{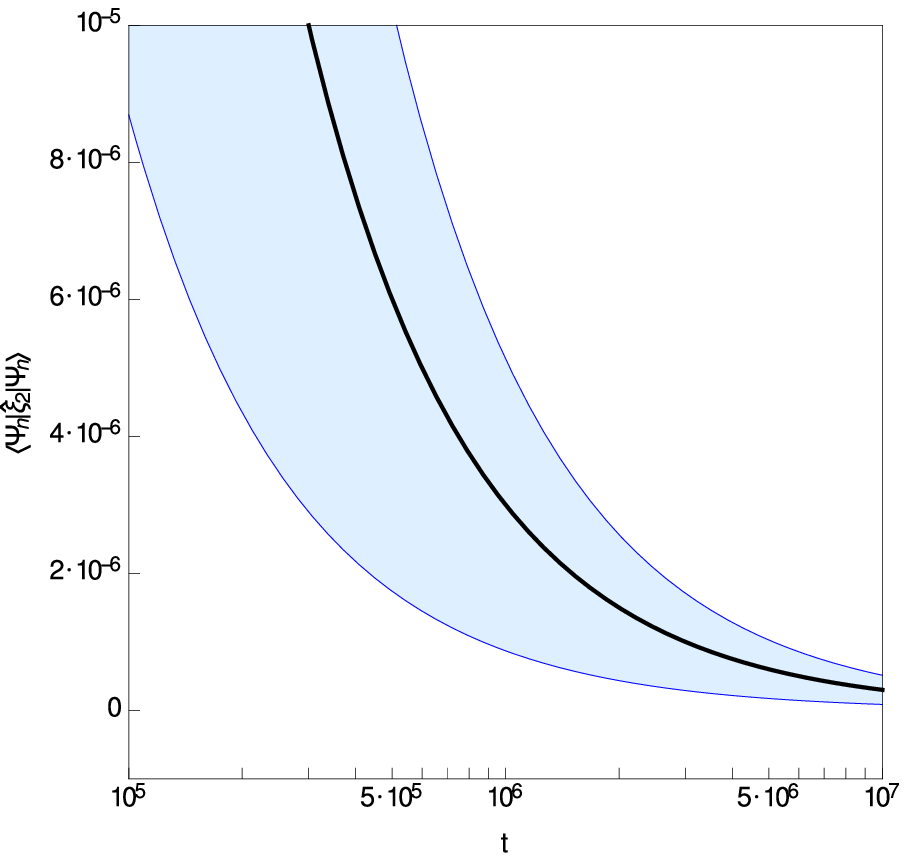}
\hspace{0.5cm}
\includegraphics[width=0.47\textwidth]{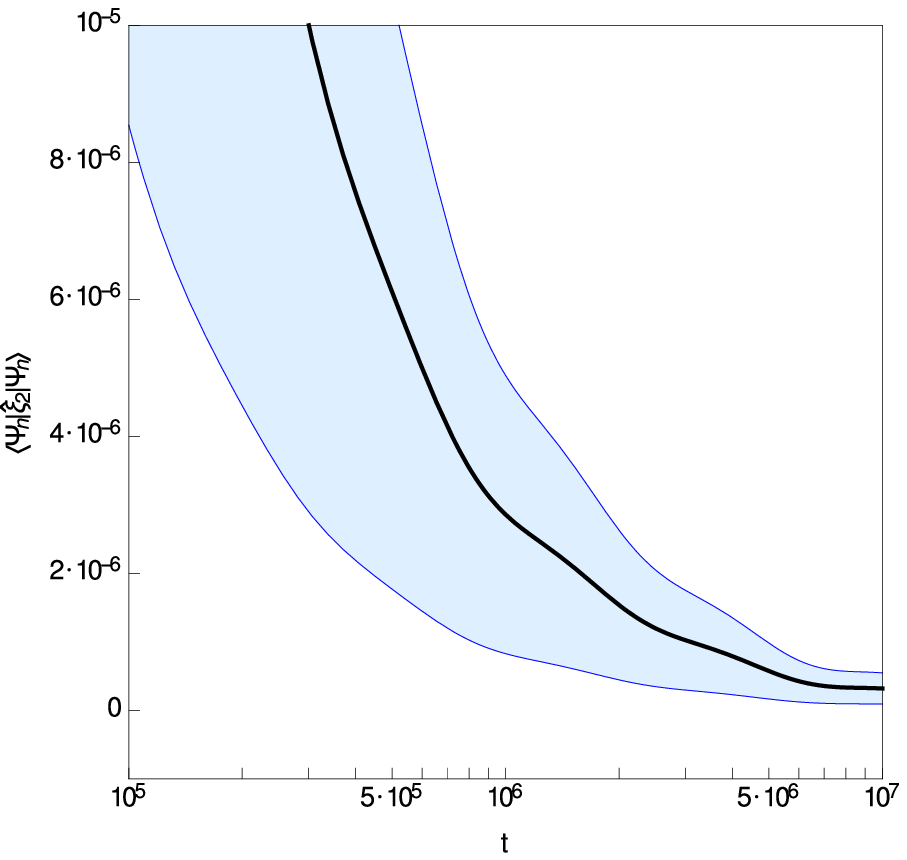}
\caption{\label{plot:2D}
The $t$ dependence of the expectation value of the operator $\hat{\xi}_2$ defined by
\eqref{eq:ExpXik1},\eqref{eq:EWavePackCond4} for  $K_1=K_2=0.01$, $K_3=0$, $\phi_1=\phi_2=0$, any $t_0<0$, $n_1=3$.
Axis of $t$ is in logarithmic scale.
The left panel corresponds to unperturbed solution
($\epsilon=0$), the right panel corresponds to perturbed solution ($\epsilon=0.01$).
The blue area defines the points for which distance from expected value is smaller than
$\sqrt{\Var{\xi_2;\Psi_n}}$ defined by \eqref{vp2} (the distance is counted along fixed $t$ line).
}
\end{center}
\end{figure}

Figure \!\ref{plot:3D} presents
the parametric curve visualizing the relative quantum perturbations. The time dependence of
the expectation values of $\hat{\xi}_2$ operator  and corresponding variances of unperturbed
and perturbed solutions are presented in Fig. \!\ref{plot:2D}.
The plots for $\hat{\xi}_4$ and $\hat{\xi}_6$ operators would look similarly so that we do not
present them.

\section{Conclusions}

It has been shown \cite{Piotr} that the perturbed classical solution to the dynamics
of the massive model of the BKL scenario exhibits the chaotic behaviour.  Our results present
the quantum dynamics corresponding to that dynamics.
Fig. \!\ref{plot:3D} shows that the relative quantum perturbations grow as the system
evolves towards the singularity, which is consistent with the corresponding
classical evolution, see Fig. 2 of \cite{Piotr}.  Since our quantum and classical perturbations
have quite  similar time evolutions, we conclude that quantization  does not destroy classical chaos.
In fact, the quantum chaos  corresponds to the classical chaos in the lowest order approximation.
Nonlinearity of  classical dynamics creates  deterministic chaos.
Non-vanishing  variances of observables of the corresponding quantum dynamics leads to stochastic
chaos.

To show that behaviour,  we have constructed the wave packets for which expectation
values of elementary observables follow the corresponding classical ones.  This choice
of quantum states leads to the scenario in which calculated expectation values of quantum
directional scale factors evolve to vanishing similarly as their classical counterparts.
However, quantum states do not represent sharp properties of a physical system.
Expectation values  are smeared quantities.  That smearing is represented
by quantum variance.

More exactly, the variance is a measure of the stochastic deviation from an
expectation value of given operator.  A quantum system is in an eigenstate of
an operator iff the variance of this operator in that state equals zero (see,
\cite{AOW} for more details).  The calculated variances depicted in
Fig. \!\ref{plot:2D} are always non-zero, which mean that the probability of
hitting the gravitational singularity is equal to zero. The non-zero variance
removes the singularity from considered quantum evolution.

The quantum randomness amplifies the deterministic classical chaos.  This
supports the hypothesis that in the region corresponding to the neighbourhood of
the classical singularity the dynamics, both classical and quantum, enters the
stochastic phase. The oscillatory behaviour of the expectation value of the
quantum scale factor increases as $t \rightarrow \infty$, which is consistent
with the classical BKL scenario \cite{BKL3,book}.

The results of the present paper support our previous results \cite{AWG,AW}
concerning the fate of the BKL singularity at the quantum level.
One of the main differences between the results of the papers \cite{AWG,AW}
and the present article is that in the former case the evolution parameter (time)
used at the quantum level was purely mathematical and was taken being equal to the classical
time by an assumption.  In the present paper, the quantum time is established by the
requirement  that the temporal and spatial variables should be treated on the same
footing at quantum level, which supports the covariance of arbitrary transformations of
these variables in general relativity.  Another important difference is quite different
implementation of the
dynamics both at classical and quantum levels. In the case \cite{AWG,AW} it was based
on Hamilton's dynamics and corresponding Schr\"{o}dinger's equation at quantum level.
Here, we quantize the solution to the classical dynamics ascribing to it corresponding
quantum system. Surprisingly, both approaches give physically  similar result:
avoiding of classical singularity at quantum level. Additionally, the approach of the present paper
addresses the issue of the fate of the classical chaos at quantum level, which was
beyond the scope of the approach used in  \cite{AWG,AW}. We have examined that issue by calculating
the variances of expectation values of quantum observables. The variances are measures of stochasticity
of considered observables at quantum level.
In fact, the calculations of variances have been ignored in the recent
quantizations of the vacuum BKL models \cite{WP1,WP2,nowa}.

The BKL conjecture states that general relativity  includes generic gravitational
singularity. Our results strongly suggest that generic singularity  can
be avoided at quantum level so that one can expect that a theory of quantum
gravity (to be constructed) has good chance to be regular.

\acknowledgments We would like to thank Vladimir Belinski and Piotr Goldstein for helpful discussions.

\appendix

\section{Essence of integral quantization}

If the  configuration space $\Pi$ is a   half-plane,

\[ \Pi := \{(p, q) \in \dR \times \dR_+\} ,~~~\dR_+ := \{x \in \dR~|~x>0 \},\]

it can be identified with the  affine group  Aff$(\dR)$ =: G.

The multiplication law can be defined as
\begin{equation}\label{A1}
(p_1, q_1) \cdot (p_2, q_2) := (p_1 + q_1 p_2, q_1 q_2)  \, .
\end{equation}
The unity of the group is $(0,1)$ and the inverse reads $(p,q)^{-1} = (-p/q, 1/q)$.

This group has two nontrivial unitary irreducible representations  realized in the Hilbert space $ L^2(\dR_+, d\nu(x)) =: \mathcal{H}$,
where  $d\nu(x)=dx/x$. We choose the one defined as follows (the second representation would lead to exactly the same results):
\begin{equation}\label{A2}
U(p,q)\psi(x)= e^{i p x} \psi(qx),~~~~\psi(x) \in \mathcal{H}~\;.
\end{equation}

Eq. \!\eqref{A2} enables defining the  continuous family of affine coherent states (ACS),
denoted $\langle x|p,q\rangle \in  \mathcal{H}$, as follows
\begin{equation}\label{A3}
 \langle x |p,q\rangle = U(p,q) \langle x | \phi\rangle \; ,
\end{equation}
where $ \langle x|\phi\rangle =: \phi (x) \in \mathcal{H}$ is the so-called   fiducial vector,
which is a free parameter (to some extent) of ACS quantization scheme.

Eq. \!\eqref{A3} can be interpreted as the correspondence
\begin{equation}\label{A4}
 (p,q) \longrightarrow |p,q\rangle \langle p,q |
\end{equation}
between  the point of the configuration space $\Pi$ and  the quantum projection  operator acting in $\mathcal{H}$.

The  irreducibility of the representation leads (due to Schur's lemma)
to the  resolution of the  unity in $L^2(\dR_+, d\nu(x))$:
\begin{equation}\label{A5}
  \frac{1}{A_\phi}\int_{G} d\mu (p,q)  |p,q\rangle \langle p,q| = \UnitOp \; ,
\end{equation}
where $d\mu (p,q) := dp\; dq/q^2$ is the left invariant measure on $G$,  and where
$A_\phi := \int_0^\infty |\phi(x)|^2 \,\frac{dx}{x^2}< \infty$ is a constant.

Using \eqref{A5} enables  quantization of almost any observable $f: \Pi \rightarrow \dR $
\begin{equation}\label{A6}
   f \longrightarrow \hat{f} = \frac{1}{ A_\phi} \int_{G} d\mu (p,q) |p,q\rangle f(p,q) \langle p,q| \, .
\end{equation}
The operator $\hat{f}: \mathcal{H} \rightarrow \mathcal{H}~$ is  symmetric  by construction.
No ordering  ambiguity occurs (notorious problem of canonical quantization). That operator is self-adjoint
if it is bounded.

For more details concerning the integral quantization, see \cite{AWT} and references therein.


\end{document}